\documentclass[aps,twocolumn,footinbib,showpacs]{revtex4-1}
\usepackage{color}
\usepackage{amsfonts,amssymb,amsmath,mathrsfs}
\usepackage{amsbsy}
\usepackage{graphicx}
\usepackage{hyperref}

\usepackage{pdfpages}
\makeatletter

\newcommand{\ssection}[1]{ \textit{#1.} }

\begin{document}

\title{Theory of magnon motive force in chiral ferromagnets}

\author{Utkan G\"ung\"ord\"u}
\email{ugungordu@unl.edu}
\affiliation{Department of Physics and Astronomy and Nebraska Center for Materials and Nanoscience, University of Nebraska, Lincoln, Nebraska 68588, USA}

\author{Alexey A. Kovalev}
\affiliation{Department of Physics and Astronomy and Nebraska Center for Materials and Nanoscience, University of Nebraska, Lincoln, Nebraska 68588, USA}


\begin{abstract}
We predict that magnon motive force can lead to temperature dependent, nonlinear chiral damping in both conducting and insulating ferromagnets. We estimate that this damping can significantly influence the motion of skyrmions and domain walls at finite temperatures. We also find that in systems with low Gilbert damping moving chiral magnetic textures and resulting magnon motive forces can induce large spin and energy currents in the transverse direction.
\end{abstract}


\maketitle

Emergent electromagnetism in the context of spintronics \cite{Zutic:RoMP2004} brings about interpretations of the spin-transfer torque \cite{Slonczewski:JMMM1996,Berger:PRB1996} and spin-motive force (SMF) \cite{Berger:1986,Volovik1987,Barnes2007,Yang:2009,Ohe:2009,Yamane:2011,Yamane.Hemmatiyan.ea:SR2014} in terms of fictitious electromagnetic fields.
In addition to providing beautiful interpretations, these concepts are also very useful in developing the fundamental understanding of magnetization dynamics. A time-dependent magnetic texture is known to induce an emergent gauge field on electrons \cite{Volovik1987}. As it turns out, the spin current generated by the resulting fictitious Lorentz force (which can also be interpreted as dynamics of Berry-phase leading to SMF) influences the magnetization dynamics in a dissipative way \cite{Volovik1987,Barnes2007,Zhang2009,Tserkovnyak2009,Wong2010,Fahnle2011,Kim2012,Kim2015a}, affecting the phenomenological Gilbert damping term in the Landau-Lifshitz-Gilbert (LLG) \cite{Gilbert2004} equation. Inadequacy of the simple Gilbert damping term  has recently been seen experimentally in domain wall creep motion \cite{Jue2015}. 
Potential applications of such studies include control of magnetic solitons such as domain walls and skyrmions \cite{Bogdanov1994,Rossler2006,Muhlbauer2009a,Yu2010,
Jonietz2011,Heinze2011,Kiselev2011,Kiselev2011,Schulz2012a,Iwasaki2013,Fert2013,Buttner2015,
Tokunaga2015,Woo2015}, 
which may lead to faster magnetic memory and data storage devices with lower power requirements \cite{Parkin2008,Sampaio.Cros.ea:NN2013,Brataas.Kent.ea:NM2012}. Recently, phenomena related to spin currents and magnetization dynamics have also been studied in the context of energy harvesting and cooling applications within the field of spincaloritronics \cite{Hatami.Bauer.ea:PRL2007,Bauer.Bretzel.ea:2010,Kovalev:PRB2009,Cahaya.Tretiakov.ea:APL2014,Kovalev:SSC2010}.

\begin{figure}[htbp]
\centering
\includegraphics[width=0.9\columnwidth]{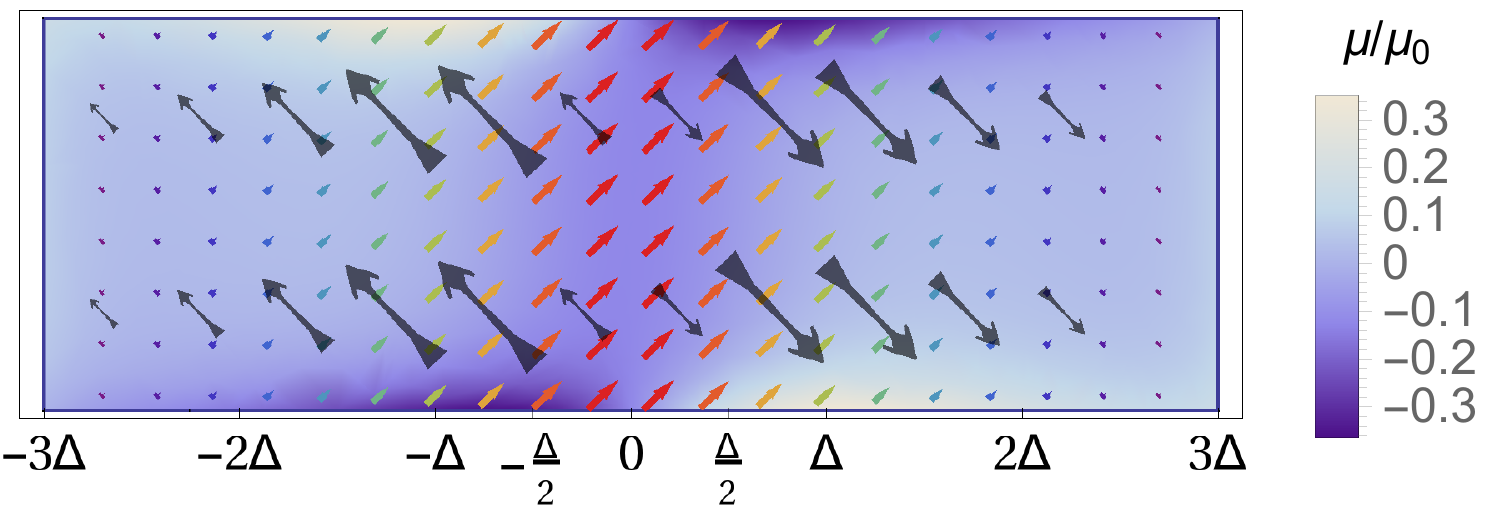}

\includegraphics[width=0.9\columnwidth]{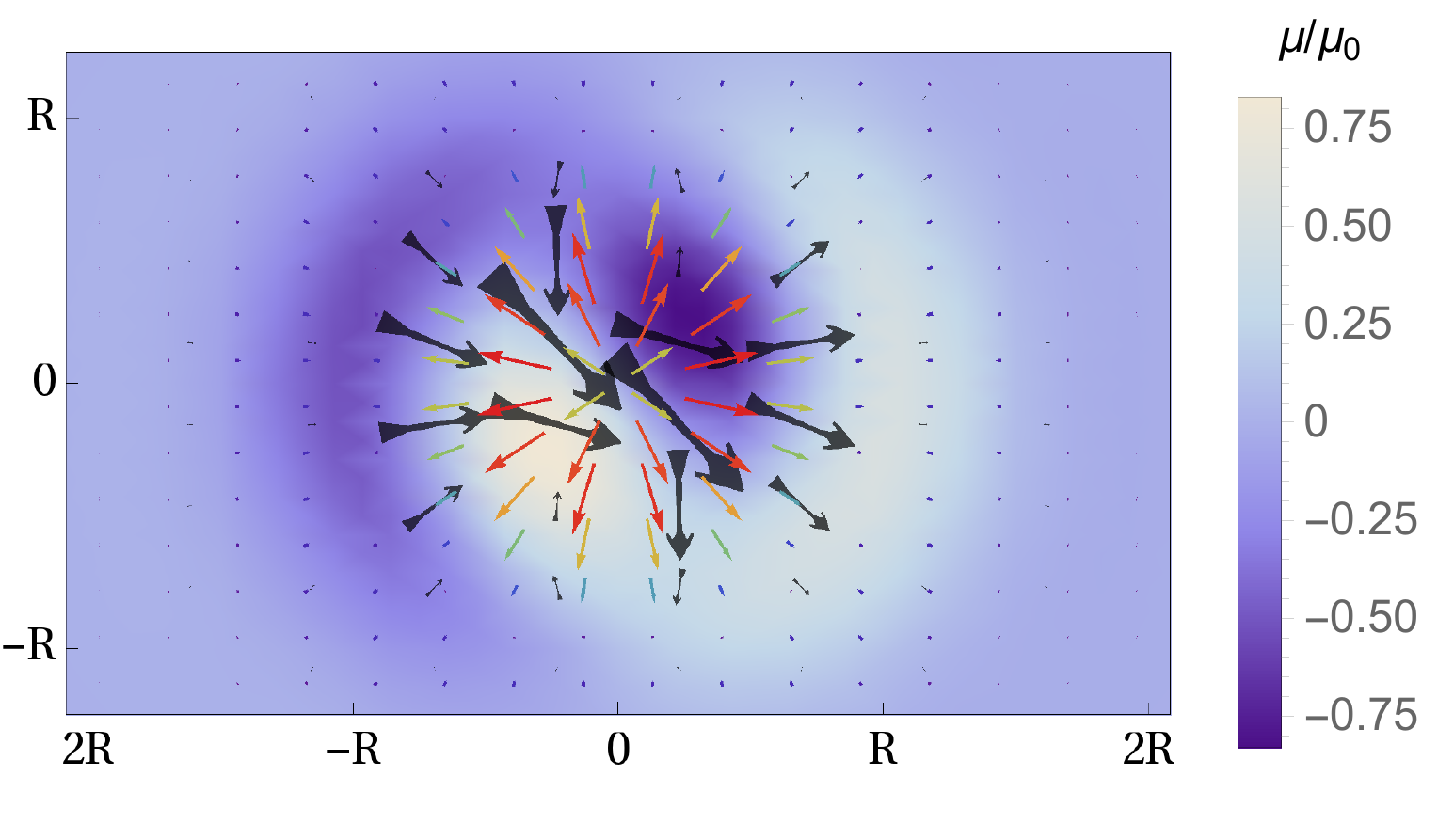}

\caption{(Color online)  A moving magnetic texture, such as a domain wall (top) or an isolated skyrmion (bottom), generates an emergent electric field and accumulates a cloud of magnons around it. In-plane component of $\boldsymbol n_s$ and electric field $\boldsymbol {\mathcal E}$ are represented by small colored arrows and large black arrows, respectively. Magnon chemical potential $\mu$  is measured in $\mu_0 = \xi \hbar v D/J\Delta$ for domain wall  and $\mu_0 = \xi \hbar v D/JR$ for skyrmion, where $\xi$ is the magnon diffusion length. Solitons are moving along $+x$ axis with velocity $v$, $n_z=\pm1$ at $x=\mp \infty$ for domain wall and $n_z=1$ at the center for the isolated hedgehog skyrmion.
Material parameters for Co/Pt ($J=16$pJ/m, $D = 4$mJ/m$^2$, $M_s=1.1$MA/m, $\alpha=0.03$, at room temperature \cite{Yang2015}) were used for domain wall leading to $\Delta\approx 7$nm, and Cu$_2$OSeO$_3$ parameters ($J  = 1.4$pJ/m, $D = 0.17$mJ/m$^2$, $s=0.5\hbar/a^3$, $a=0.5$nm with $\alpha=0.01$, at $T\sim 50$K \cite{Kovalev2015a}) for skyrmion leading to $R\approx50$nm. System size is taken to be $6\Delta \times 2\Delta$ for domain wall and $6R \times 6R$ for skyrmion.
 }
\label{fig:diffusion}
\end{figure}

Magnons, the quantized spin-waves in a magnet, are present in both conducting magnets and insulating magnets. Treatment of spin-waves with short wavelengths as quasiparticles allows us to draw analogies from systems with charge carriers. For instance, the flow of thermal magnons generates a spin transfer torque (STT) \cite{Kovalev2012,Kovalev2014a,Kim2015} and a time-dependent magnetic texture exerts a magnon motive force.
According to the Schr\"odinger-like equation which governs the dynamics of magnons in the adiabatic limit \cite{Kovalev2014a}, the emergent ``electric" field induced by the time-dependent background magnetic texture exerts a ``Lorentz force" on magnons, which in turn generates a current by ``Ohm's law" (see Fig.~\ref{fig:diffusion}).
Despite the similarities, however, the strength of this feedback current has important differences from its electronic analog: it is inversely proportional to the Gilbert damping and grows with temperature.

In this paper, we formulate a theory of magnon feedback damping induced by the magnon motive force. We find that this additional damping strongly affects the dynamics of magnetic solitons, such as domain walls and skyrmions, in systems with strong Dzyaloshinskii-Moriya interactions (DMI). We also find that the magnon motive force can lead to magnon accumulation (see Fig.~\ref{fig:diffusion}), non-vanishing magnon chemical potential, and large spin and energy currents in systems with low Gilbert damping. To demonstrate this, we assume diffusive transport of magnons in which the magnon non-conserving relaxation time, $\tau_\alpha$, is larger compared to the magnon conserving one, $\tau_m$ ($\tau_\alpha \gg \tau_m$). For the four-magnon thermalization, $\tau_m=\hbar/(k_B T)(T_c/T)^3$, and for LLG damping, $\tau_\alpha=\hbar/\alpha k_B T$, this leads to the constraint $\alpha(T_c/T)^3 \ll 1$ \cite{Dyson:1956,Bender:2014}.

\ssection{Emergent electromagnetism for magnons}We initially assume that the magnon chemical potential is zero. The validity of this assumption is confirmed in the last section. Effects related  to emergent electromagnetism for magnons can be captured by considering a ferromagnet well below the Curie temperature. We use the stochastic LLG equation:
\begin{align}
s(1+\alpha \boldsymbol n \times) \dot {\boldsymbol n}  =\boldsymbol n \times (\boldsymbol H_\text{eff} + \boldsymbol h), \label{eq:LLG0}
\end{align}
where $s$ is the spin density along $\boldsymbol n$, $\boldsymbol H_\text{eff} = -\delta_{\boldsymbol n} F[\boldsymbol n]$ is the effective magnetic field, $F[\boldsymbol n] = \int d^3 r \mathcal F(\boldsymbol n)$ is the free energy
 and $\boldsymbol h$ is the random Langevin field. It is convenient to consider the free energy density $\mathcal F(\boldsymbol n) = J  (\partial_i \boldsymbol n)^2/2 + \hat D \boldsymbol e_i \cdot (\boldsymbol n \times \partial_i \boldsymbol n) + \boldsymbol H \cdot \boldsymbol n + K_u n_z^2$ where $J$ is the exchange coupling, $\hat D$ is a tensor which describes the DMI \footnote{The DM tensor $\hat D$ represents a general form of DMI. In particular, bulk inversion asymmetry contributes to $\hat D$ as $D_0 \openone$ whereas structure inversion asymmetry contributes as $D_R \boldsymbol e_z \times$.}, $\boldsymbol H = M_s \mu_0 H_a \boldsymbol e_z$ describes the magnetic field, $K_u$ denotes the strength of uniaxial anisotropy, $M_s$ is the saturation magnetization, $H_a$ is the applied magnetic field, and summation over repeated indices is implied.
At sufficiently high temperatures, the form of anisotropies is unimportant for the discussion of thermal magnons and can include additional magnetostatic and magnetocrystalline contributions. 
  
Linearized dynamics of magnons  can be captured by the following equation \cite{Gungordu2016a}:
\begin{align}
 s (i\partial_t+\boldsymbol n_s \cdot \boldsymbol A_t) \psi =& \left[J (\partial_i/i -\boldsymbol n_s \cdot [\boldsymbol A_i  -\boldsymbol D_i/J])^2  + \varphi  \right]\psi,
\label{eq:LLG-fast-dynamics}
\end{align}
where  $\varphi$ absorbs effect of anisotropies, DMI and the magnetic field, $\psi = \boldsymbol n_f\cdot(\boldsymbol e'_x + i \boldsymbol e'_y)$ describes fluctuations $ \boldsymbol  n_f=\boldsymbol n - \boldsymbol n_s\sqrt{1- \boldsymbol n_f^2} $ around slow component $\boldsymbol n_s$ ($|\boldsymbol n| = |\boldsymbol n_s|=1$, $\boldsymbol n_s \perp \boldsymbol n_f$) in a rotated frame in which $\boldsymbol e'_z=\boldsymbol n_s$,  $\boldsymbol D_i=\hat D \boldsymbol e_i$, and $\boldsymbol A_\mu\times \equiv \hat R \partial_\mu \hat R^T$ corresponds to the gauge potential with  $\mu=x,y,z,t$. Note that in the rotated frame,
we have $\boldsymbol n\rightarrow\boldsymbol n'=\hat{R}\boldsymbol n$ and
$\partial_{\mu}\rightarrow(\partial_{\mu}-\boldsymbol A'_\mu\times)$
with $\boldsymbol A'_\mu\times=(\partial_{\mu}\hat{R})\hat{R}^{T}$. 
In deriving Eq.~\ref{eq:LLG-fast-dynamics}, we assumed that the exchange interaction is the dominant contribution and neglected the coupling between the circular components of $\psi$ and $\psi^\dagger$ due to anisotropies \cite{Dugaev2005,Kovalev2014a}.

The gauge potential in Eq.~(\ref{eq:LLG-fast-dynamics}) leads to a reactive torque in the LLG equation for the slow dynamics \cite{Tatara2015}. Alternatively, one can simply average Eq.~(\ref{eq:LLG0}) over the fast oscillations arriving at the LLG equation with the magnon torque term \cite{Kovalev2012}:
\begin{align}
 s(1+\alpha \boldsymbol n_s \times) \dot {\boldsymbol n}_s -  \boldsymbol n_s \times \boldsymbol H_\text{eff}^s = \hbar(\boldsymbol j \cdot \boldsymbol {\mathcal D}) \boldsymbol n_s,
\label{eq:STT}
\end{align}
where $\boldsymbol H_\text{eff}^s = -\delta_{\boldsymbol n_s} F[\boldsymbol n_s]$ is the effective field for the slow magnetization calculated at zero temperature \footnote{We disregard \unexpanded{$\mathcal O( \langle\boldsymbol n_f^2 \rangle )$} corrections assuming temperatures well below the Curie temperature. These corrections can be readily reintroduced.}, $j_i = (J/\hbar)\langle \boldsymbol n_s \cdot (\boldsymbol n_f \times \partial_i \boldsymbol n_f) \rangle$ is the magnon current and
$\mathcal D_i = \partial_i +  (\hat D \boldsymbol e_i/J) \times$ is the chiral derivative \cite{Kim2013a,Tserkovnyak2014,Gungordu2016a}.

\ssection{Magnon feedback damping}The magnon current $\boldsymbol j$ is induced in response to the emergent electromagnetic potential, and can be related to the driving electric field $\mathcal E_i = \hbar \boldsymbol n_s \cdot (\partial_t \boldsymbol n_s \times \mathcal D_i \boldsymbol n_s)$ by local Ohm's law $\boldsymbol j = \sigma \boldsymbol {\mathcal E}$ where $\sigma$ is magnon conductivity.  The induced electric field $\boldsymbol {\mathcal E}$ can be interpreted as a magnon generalization of the spin motive force \cite{Barnes2007}. The magnon feedback torque $\boldsymbol \tau = \hbar \sigma  (\boldsymbol {\mathcal E} \cdot \boldsymbol {\mathcal D}) \boldsymbol n_s$
has dissipative effect on magnetization dynamics and leads to a damping tensor $\hat \alpha_\text{emf} =\eta (\boldsymbol n_s \times \mathcal D_i \boldsymbol n_s) \otimes (\boldsymbol n_s \times \mathcal D_i \boldsymbol n_s)$ in the LLG equation with $ \eta = \hbar^2 \sigma/ s$ \footnote{In a quasi two-dimensional or two-dimensional system, surface spin density should be used which can be obtained from the bulk spin density as $ s_\text{2D} = s_\text{3D} t$ where $t$ is the layer thickness.}. 

A general form of the feedback damping should also include the contribution from the dissipative torque \cite{Kovalev2012}. Here we introduce such $\beta$-terms phenomenologically which leads to the LLG equation:
\begin{equation}
s (1 +  \boldsymbol n_s \times  [\hat\alpha+\hat \Gamma]) \dot {\boldsymbol n}_s - \boldsymbol n_s \times \boldsymbol H^s_\text{eff} = \boldsymbol \tau, \label{eq:LLG-damping}
\end{equation}
where $\boldsymbol \tau$ is the magnon torque term and we separated the dissipative $\hat\alpha$ and reactive $\hat\Gamma$ contributions:
\begin{align}
&\hat \alpha =   \alpha \boldsymbol  + \hat\alpha_\text{emf} - \eta  \beta^2 \mathcal D_i \boldsymbol n_s \otimes \mathcal D_i \boldsymbol n_s,  \\
&\hat \Gamma =   \eta \beta[(\boldsymbol n_s \times \mathcal D_i \boldsymbol n_s) \otimes \mathcal D_i \boldsymbol n_s - \mathcal D_i \boldsymbol n_s \otimes (\boldsymbol n_s \times \mathcal D_i \boldsymbol n_s)],\nonumber
\end{align}
where in general the form of chiral derivatives in the $\beta$-terms can be different. Given that $\beta$ and $\alpha$ are typically small for magnon systems, the term $\hat\alpha_\text{emf}$ will dominate the feedback damping tensor. An unusual feature of the chiral part of the damping is that it will be present even for a uniform texture. While the DMI prefers twisted magnetic structures, this can be relevant in the presence of an external magnetic field strong enough to drive the system into the ferromagnetic phase.
 
In conducting ferromagnets, charge currents also lead to a damping tensor of the same form where the strength of the damping is characterized by $\eta_e = \hbar^2 \sigma_e/4 e^2 s$ with $\sigma_e$ as the electronic conductivity \cite{Zhang2009,Wong2010,Kim2012}, which should be compared to $\eta$ in  conducting ferromagnets where both effects are present.
Since the magnon feedback damping $\eta$ grows as $\propto 1/\alpha$, the overall strength of magnon contribution can quickly become dominant contribution in ferromagnets with small Gilbert damping. Under the assumption that magnon scattering is dominated by the Gilbert damping such that the relaxation time is given by $\tau_\alpha = 1/2\alpha \omega$, magnon conductivity is  given by $\sigma_\text{3D} \sim 1/6 \pi^2 \lambda \hbar \alpha$ in three-dimensions and $\sigma_\text{2D} \sim 1/4 \pi \hbar \alpha$ in two-dimensions \cite{Kovalev2012} where $\lambda=\sqrt{\hbar J/s k_B T}$ is the wavelength of the thermal magnons.
For Cu$_2$OSeO$_3$ in ferromagnetic phase, we find $\eta \approx 2$nm$^{2}$. Similarly, for a Pt/Co/AlO$_x$ thin film of thickness $t=0.6$nm yield $\eta\approx 1$nm$^2$ at room temperature. 
This shows that the magnon feedback damping can become significant in ferromagnets with sharp textures and strong DMI.

\ssection{Domain wall dynamics}We describe the domain wall profile in a ferromagnet with DMI  by Walker ansatz  $\tan(\theta(x,t)/2) = \exp(\pm [x - X(t)]/\Delta)$ where  $X(t)$ and $\phi(t)$ denote the center position and tilting angle of the domain wall \footnote{While a rigorous analysis of domain wall dynamics in the presence of a strong DMI should take domain wall tilting into account in general, in the particular case of a domain wall driven by a perpendicular field, the $(X,\phi)$ model remains moderately accurate for studying the effects of feedback damping \cite{Boulle2013,Kim2015a}.}, $\Delta = \sqrt{J/K_0}$ is the domain wall width, $K_0 = K_u - \mu_0 M_s^2/2$ includes the contributions from uniaxial anisotropy as well as the demagnetizing field and $\hat D = -D (\sin\gamma \openone + \cos\gamma \boldsymbol e_z \times)$ contains DMI due to bulk and structure inversion asymmetries whose relative strength is determined by $\gamma$. After integrating the LLG equation, we obtain the equations of motion for a domain wall driven by external perpendicular field \cite{Thiaville2012}:
\begin{align}
\Gamma_{XX} \dot X/\Delta + \dot \phi = F_X ,\quad \Gamma_{\phi \phi}\dot \phi - \dot X/\Delta  = F_\phi ,
\end{align}
where $\Gamma_{XX} = \alpha + \eta (D/J)^2 \sin^2(\gamma+\phi)/3$ and $\Gamma_{\phi \phi} = \alpha + \eta[2 /3 \Delta^2  + (\pi D/2J \Delta) \cos(\gamma+\phi) + (D/J)^2 \cos^2(\gamma+\phi) ]$ are dimensionless angle-dependent drag coefficients, $F_X = H/s$ and $F_\phi = [K \sin2\phi + \sin(\gamma+\phi) D\pi /2 \Delta]/s$ are generalized ``forces" associated with the collective coordinates $X$ and $\phi$, $K$ is the strength of an added anisotropy corresponding, e.g., to  magnetostatic anisotropy  $K=N_x \mu_0 M_s^2/2$ where $N_x$ is the demagnetization coefficient. In deriving these equations, we have neglected higher order terms in $\alpha$ and $\beta$ \footnote{This equation is similar to the equation obtained for electronic feedback damping in \cite{Kim2015a} but for magnons $\Gamma_{XX}$ and $\Gamma_{\phi\phi}$ are determined by $D/J$ rather than $\tilde \alpha_R$ (a parameter which is taken to be independent from $D$), which leads to different conclusions. In \cite{Kim2015a}, the chiral derivative associated with SMF is parameterized by $\tilde\alpha_R$ as $\mathcal D_i = \partial_i + (\tilde\alpha_R \boldsymbol e_z \times)\boldsymbol e_i \times$. For magnons, $D/J$ corresponds to $\tilde\alpha_R$.}.

\begin{figure}[!htb]
\centering
\includegraphics[width=0.48\columnwidth]{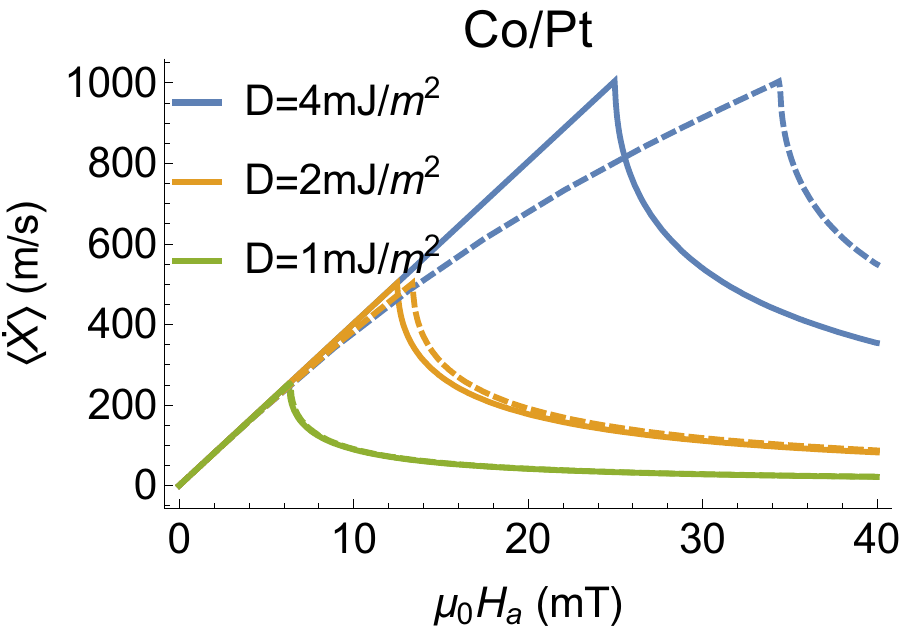}
\includegraphics[width=0.48\columnwidth]{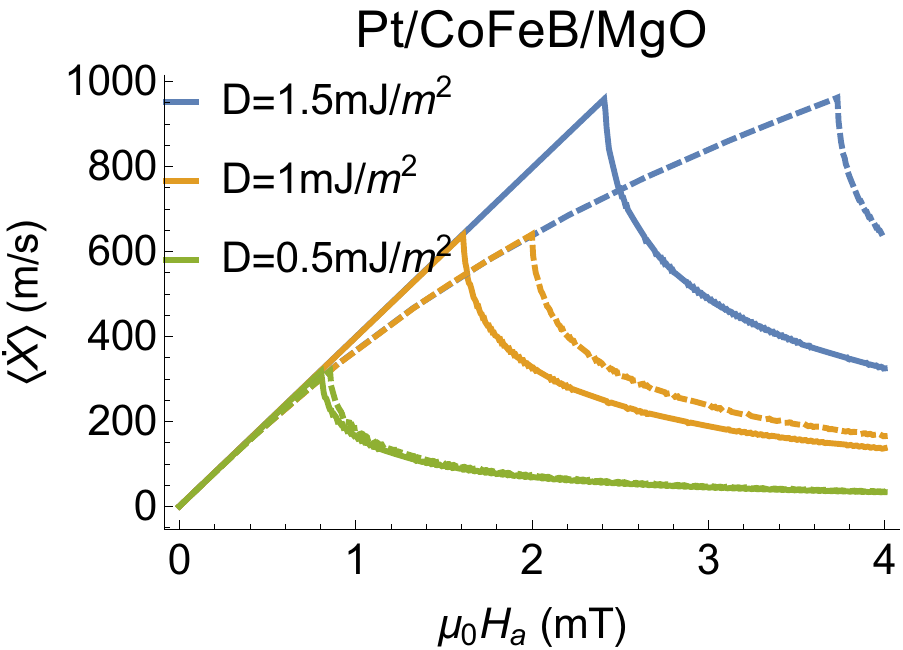}
\caption{(Color online) Domain wall velocity as a function of the magnetic field and varying strength of DMI for Co/Pt and Pt/CoFeB/MgO films. Solid (dashed) lines correspond to dynamics at zero (room) temperature. We used material parameters $M_s=1.1$MA/m, $J=16$pJ/m, $K_0 = 0.34$MJ/m$^3$, $\alpha=0.03$ \cite{Yang2015} for Co/Pt, and $M_s=0.43$MA/m, $J=31$pJ/m, $K_0 = 0.38$MJ/m$^3$, $\alpha = 4 \times 10^{-3}$ \cite{Woo2015,Yamanouchi2011,Liu2011} for Pt/CoFeB/MgO.
 }
\label{fig:dw-plot}
\end{figure}

Time-averaged domain wall velocity obtained from numerical integration of the equations of motion for a Co/Pt interface with Rashba-like DMI is shown in Fig.~\ref{fig:dw-plot}.
Thermal magnon wavelength at room temperature ($\approx 0.3$nm) is much shorter than the domain wall size $\Delta=\sqrt{J/K_0} \approx 7$nm, so the quasiparticle treatment of magnons is justified.
We observe that damping reduces the speed at fixed magnetic field, and this effect is enhanced with increasing DMI strength $D$ and diminishing the Gilbert damping $\alpha$ (see Fig.~\ref{fig:dw-plot}).

Another important observation is that in the presence of the feedback damping, the relation between applied field and average domain wall velocity becomes nonlinear. This is readily seen from steady state solution of the equations of motion before the Walker breakdown with $\phi=\phi_0$ which solves $\sin(\gamma + \phi_0) D\pi/2  s \Delta = -[H/s][\alpha + \eta (D/J)^2 \sin^2(\gamma+\phi_0)/3]^{-1}$ (noting that $D/\Delta \gg K$, implying a N\'eel domain wall \cite{Thiaville2012,Buijnsters2016}) and $X = v t$, leading to the cubic velocity-field relation $(s v/\Delta)(\alpha + \eta[2  s v /J\pi]^2/3) = H$ for field-driven domain wall motion. The angle $\phi_0$ also determines the tilting of $\boldsymbol {\mathcal E}$ as seen in Fig. \ref{fig:diffusion}.

\ssection{Skyrmion dynamics}Under the assumption that the skyrmion retains its internal structure as it moves, we treat it as a magnetic texture $\boldsymbol n_s = \boldsymbol n_s(\boldsymbol r - \boldsymbol q(t))$ with $\boldsymbol q(t)$ being the time-dependent position (collective coordinate \cite{Thiele1973}) of the skyrmion. 
We consider the motion of a skyrmion under the temperature gradient $\boldsymbol \nabla \chi=-\boldsymbol \nabla T/T$, which exerts a magnon torque: \begin{equation}
\boldsymbol \tau =(1+\beta_T \boldsymbol n_s \times)(L \boldsymbol \nabla \chi \cdot  \boldsymbol {\mathcal D}) \boldsymbol n_s,
\end{equation}
where $L$ is the spin Seebeck coefficient and $\beta_T$ is the ``$\beta$-type" correction. These are given by $L_\text{3D} \sim k_B T/6 \pi^2 \lambda \alpha$ and $\beta_T\approx 3 \alpha/2$ in three-dimensions and $L_\text{2D} \sim k_B T/4 \pi  \alpha$ and $\beta_T\approx\alpha$ in two-dimensions within the relaxation time approximation \cite{Kovalev2014a,Kim2015}.
 Multiplying the LLG equation Eq.~(\ref{eq:LLG-damping}) with $\int d^2 r \partial_{q_j} \boldsymbol n_s \cdot \boldsymbol n_s \times$ and substituting $\dot {\boldsymbol n}_s = -\dot q_i \partial_{q_i} \boldsymbol n_s$, we obtain the equation of motion for $\boldsymbol v = \dot{\boldsymbol q}$:
\begin{align}
s\left(W -Q \boldsymbol z \times    \right)\boldsymbol v  +
\left(\beta_T \eta_D -Q \boldsymbol z \times  \right)L \boldsymbol \nabla \chi = \boldsymbol F.
\label{eq:Thiele}
\end{align}
Above, $W = \eta_0 \alpha + \eta \alpha_0$ can be interpreted as the contribution of the renormalized Gilbert damping,
 $Q =  \int d^2 r \boldsymbol n_s \cdot (\partial_x \boldsymbol n_s \times \partial_y \boldsymbol n_s)/4\pi$ is the topological charge of the skyrmion, $\eta_0$ is the dyadic tensor, $\eta_D$ is the chiral dyadic tensor which is $\sim \eta_0$ for isolated skyrmions and vanishes for skyrmions in SkX lattice \cite{Gungordu2016a} (detailed definitions of these coefficients are given in the Supplemental Material \footnote{See Supplemental Material at the end.}). The ``force" term $\boldsymbol F=-\boldsymbol\nabla U(\boldsymbol q)$ due to the effective skyrmion potential $U(\boldsymbol q)$ is relevant for systems with spatially-dependent anisotropies \cite{Yu2016}, DMI \cite{Diaz2015}, or magnetic fields.
In deriving this equation, we only considered the dominant feedback damping contribution $\hat\alpha_\text{emf}$ which is justified for small $\alpha$ and $\beta$. For temperature gradients and forces along the $x$-axis we obtain velocities:
\begin{align}
v_x =&  \frac{-L  \partial_x \chi(Q^2 + W \beta_T \eta_D)  + F_x W}{s(Q^2 + W^2)},  \nonumber \\
v_y =&  \frac{-L  \partial_x \chi Q(\beta_T \eta_D - W)+ F_x Q}{s(Q^2 + W^2)}.
\label{eq:v}
\end{align}
The Hall angle defined as $\tan \theta_H = v_y/v_x$ is strongly affected by the renormalization of $W$ since $\tan \theta_H =Q/W$ for a ``force" driven skyrmion and $\tan \theta_H \approx(\beta_T \eta_0 -  W)/Q$ for a temperature gradient driven skyrmion. Similar to the domain wall velocity in Fig.~\ref{fig:dw-plot}, the Hall effect will depend on the overall temperature of the system. We find that for a skyrmion driven by $\partial_x \chi$, the Hall angle  $\theta_H$  may flip the sign in magnets with strong DMI as the temperature increases. 
We estimate this should happen in Cu$_2$OSeO$_3$ at $T\sim 50$K using a typical radial profile for a rotationally symmetric skyrmion given by Usov ansatz $\cos(\theta/2) = (R^2 - r^2)/(R^2 + r^2)$ for $r \leq R$ and $R\approx 2\pi J/D \approx 52$nm.

\ssection{Magnon pumping and accumulation}The motion of skyrmions induces a transverse magnon current across the sample. This effect can be quantified by the average magnon current due to magnon motive force per skyrmion: 
\begin{equation}
\boldsymbol j = \sigma  \int d^2 r \boldsymbol {\mathcal E}/\pi R^2 =(\boldsymbol v \times \boldsymbol e_z) 4 \sigma \hbar^2 Q /R^2.\label{eq:MMC}
\end{equation} 
The current can only propagate over the magnon diffusion length; thus, it can be observed in materials with large magnon diffusion length or small Gilbert damping.

\begin{figure}[htbp]
\centering
\includegraphics[width=1\columnwidth]{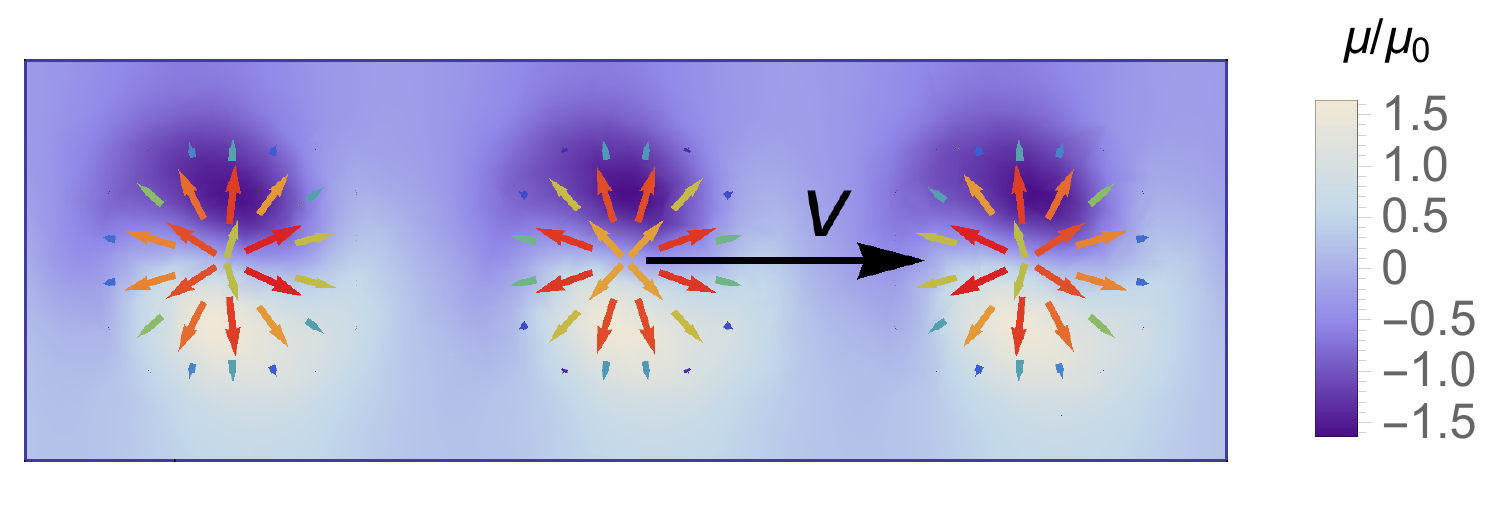}

\caption{(Color online) An array of moving skyrmions (only 3 shown in the figure) induces a transverse current and accumulation of magnons along the edges . $\mu$ is obtained by numerically solving the diffusion equation using material parameters for Pt/CoFeB/MgO given in the caption of Fig. \ref{fig:dw-plot} with $R=35$nm. System height and distance between skyrmion centers are taken to be $3R$.
}
\label{fig:skyrmions-diffusion}
\end{figure}

So far, we have assumed a highly compressible limit in which we disregard  any build up of the magnon chemical potential $\mu$. In a more realistic situation the build up of the chemical potential will lead to magnon diffusion. To illustrate the essential physics, we consider a situation in which the temperature is uniform. For slow magnetization dynamics in which magnons quickly establish a stationary state (i.e. $R/v \gg \tau_\alpha$ for skyrmions and $\Delta/\dot X \gg \tau_\alpha$ for domain walls, which is satisfied at high enough temperatures) we write a stationary magnon diffusion equation: 
\begin{align}
\nabla^2 \mu =& \frac{\mu}{\xi^2}  + \boldsymbol \nabla \cdot \boldsymbol{\mathcal E},\label{eq:diffusion}
\end{align}
where $\xi=\lambda/2\pi\alpha$ is the magnon diffusion length and we used the local Ohm's law $-\boldsymbol \nabla \mu =  \boldsymbol j/\sigma - \boldsymbol{\mathcal E}$. Renormalization of magnon current in Eq.~(\ref{eq:STT}) then follows from solution of the screened Poisson equation $\boldsymbol j= \sigma\boldsymbol{\mathcal E}+ (\sigma/4\pi)\boldsymbol\nabla\int d^3 r' (\boldsymbol \nabla' \cdot \boldsymbol{\mathcal E}) e^{-|\boldsymbol r-\boldsymbol r'|/\xi} /|\boldsymbol r-\boldsymbol r'|$ in three dimensions and $\boldsymbol j= \sigma\boldsymbol{\mathcal E}+ (\sigma/2\pi)\boldsymbol\nabla\int d^2 r' (\boldsymbol \nabla' \cdot \boldsymbol{\mathcal E}) K_0(|\boldsymbol r-\boldsymbol r'|/\xi)$ in two dimensions where $K_0$ is the modified Bessel function for an infinitely large system \footnote{As such, dynamics of magnetic solitons depend on temperature through $\sigma$ and $\xi$.}. By analyzing the magnon current due to magnon accumulation analytically and numerically, we find that renormalization becomes important when the  length associated with the magnetic texture is much smaller than the magnon diffusion length.

Finally, we numerically solve Eq.~(\ref{eq:diffusion}) for isolated solitons (see Fig.~\ref{fig:diffusion}) and for an array of moving skyrmions (see Fig.~\ref{fig:skyrmions-diffusion}). Given that the width of the strip in Fig.~\ref{fig:skyrmions-diffusion} is comparable to the magnon diffusion length one can have substantial accumulation of magnons close to the boundary. Spin currents comparable to the estimate in Eq.~(\ref{eq:MMC}) can be generated in this setup and further detected by the inverse spin Hall effect \cite{Weiler:Prl2013}. From Eq.~(\ref{eq:MMC}), for a skyrmion with $R=35$nm moving at $10$m/s in Pt/CoFeB/MgO with $D=1.5$mJ/m$^2$ \cite{Woo2015}, we obtain an estimate for spin current $j_s=j \hbar \sim 10^{-7}$ J/m$^2$ which roughly agrees with the numerical results. This spin current will also carry energy and as a result will lead to a temperature drop between the edges.

\ssection{Conclusion}We have developed a theory of magnon motive force in chiral conducting and insulating ferromagnets. The magnon motive force leads to temperature dependent, chiral feedback damping. The effect of this damping can be seen in the non-linear, temperature dependent behavior of the domain wall velocity. In addition, observation of the temperature dependent Hall angle of skyrmion motion can also reveal this additional damping contribution. We have numerically confirmed the presence of the magnon feedback damping in finite-temperature micromagnetic simulations of Eq.~(\ref{eq:LLG0}) using MuMaX3 \cite{Vansteenkiste2014}.

Magnon pumping and accumulation will also result from the magnon motive force. Substantial spin and energy currents can be pumped by a moving chiral texture in systems in which the size of magnetic textures is smaller or comparable  to the magnon diffusion length. Further studies could concentrate on magnetic systems with low Gilbert damping, such as yttrium iron garnet (YIG), in which topologically non-trivial bubbles can be realized.  

This work was supported primarily by the DOE Early Career Award DE-SC0014189, and in part by the NSF under Grants Nos. Phy-1415600, and DMR-1420645 (UG).

\bibliographystyle{apsrev}
\bibliography{skyrmion,extra,MyBIB}



\section*{Supplementary material}

# Supplemental Material for
# Theory of Magnon Motive Force in Chiral Ferromagnets

Utkan Güngördü and Alexey A. Kovalev

*Department of Physics and Astronomy and Nebraska Center for Materials and Nanoscience,*
*University of Nebraska, Lincoln, Nebraska 68588, USA*

## THIELE'S EQUATION OF MOTION FOR SKYRMION

We consider the motion of a rotationally symmetric skyrmion under the influence of applied temperature gradient. Assuming that skyrmion drifts without any changes to its internal structure, $\boldsymbol{n}_s(\boldsymbol{r},t) = \boldsymbol{n}_s(\boldsymbol{r} - \boldsymbol{q}(t))$ where $\boldsymbol{q}$ is the position of the skyrmion, we multiply the LLG equation with the operator $\int d^2 r \partial_{q_j} \boldsymbol{n}_s \cdot \boldsymbol{n}_s \times$ and integrate over the region containing the skyrmion and obtain the following equation motion:

$$s(W - Q'\boldsymbol{z}\times)\boldsymbol{v} + (\beta_T \eta_D - Q\boldsymbol{z}\times) L \boldsymbol{\partial}\chi = \boldsymbol{F}. \tag{1}$$

where $\boldsymbol{v} = \dot{\boldsymbol{q}}$ is the skyrmion velocity, $W = \eta_0 \alpha + \eta(\alpha_0 - \beta^2 \alpha_2)$, $L$ is the spin Seebeck coefficient, $\chi = 1/T$, $\boldsymbol{F} = -\boldsymbol{\nabla} U$, $Q$ is the topological charge defined in the main text and $Q' = Q - \eta\beta\alpha_1$. In terms of the polar coordinates $(\theta, \phi)$ of $\boldsymbol{n}_s$, the dyadic tensor $\eta_0$ and the chiral dyadic tensor $\eta_D$ are given by

$$\eta_0 = \pi \int_0^R dr \left( \frac{(r\partial_r \theta)^2 + \sin^2 \theta}{r} \right), \qquad \eta_D = \eta_0 + \frac{D}{J}\pi \int_0^R dr (\sin\theta \cos\theta + r\partial_r \theta), \tag{2}$$

The damping terms $\alpha_i$ can be expanded in powers of $D/J$ as $\alpha_i = \sum_j \alpha_{\beta^i, D^j}(D/J)^j$, where $\alpha_{\beta^i, D^j}$ is given by

$$\begin{aligned}
\alpha_{\beta^0, D^2} &= \pi \int_0^R dr \frac{(r\partial_r \theta)^2 \cos^2 \theta + \sin^2 \theta}{r} \\
\alpha_{\beta^0, D} &= \pi \int_0^R dr (r\partial_r \theta) \frac{r\partial_r \theta \tan\theta \cos^2 \theta + \sin^2 \theta}{r^2} \\
\alpha_{\beta^0, D^0} &= \pi \int_0^R dr (r\partial_r \theta)^2 \frac{2\sin^2 \theta}{r^3}
\end{aligned} \tag{3}$$

$$\begin{aligned}
\alpha_{\beta, D^2} &= 2\pi \int_0^R dr (\partial_r \theta) \sin\theta (\cos^2 \theta + 1) \\
\alpha_{\beta, D} &= 4\pi \int_0^R dr (\partial_r \theta) \sin\theta \frac{\cos^2 \theta \tan\theta + r\partial_r \theta}{r} \\
\alpha_{\beta, D^0} &= 2\pi \int_0^R dr (\partial_r \theta) \sin\theta \frac{\cos^2 \theta \tan^2 \theta + (r\partial_r \theta)^2}{r^2}
\end{aligned} \tag{4}$$

$$\begin{aligned}
\alpha_{\beta^2, D^2} &= \pi \int_0^R dr \left( \frac{\cos^2 \theta \sin^2 \theta + (r\partial_r \theta)^2}{r} \right) \\
\alpha_{\beta^2, D} &= 2\pi \int_0^R dr \left( \frac{\cos^2 \theta \sin^2 \theta \tan\theta + (r\partial_r \theta)^3}{r^2} \right) \\
\alpha_{\beta^2, D^0} &= \pi \int_0^R dr \left( \frac{\cos^2 \theta \sin^2 \theta \tan^2 \theta + (r\partial_r \theta)^4}{r^3} \right)
\end{aligned} \tag{5}$$

The integrals can be evaluated by using an approximate radial profile for $\theta(r)$. Using Usov ansatz yields the values enumerated in Table I for $\alpha_{\beta^i, D^j}$. Only the dominant terms are kept in the main text.

|   | 1 | $D/J$ | $(D/J)^2$ |
|---|---|---|---|
| 1 | $\frac{496\pi}{15R^2}$ | $\frac{52\pi}{5R}$ | $\frac{16\pi}{5}$ |
| $\beta$ | $\frac{C}{R^2}$ | $\frac{472\pi}{15R}$ | $\frac{16\pi}{3}$ |
| $\beta^2$ | $\frac{5056}{105R^2}$ | $\frac{2804\pi}{105R}$ | $\frac{464\pi}{105}$ |

TABLE I. List of feedback damping coefficients $\alpha_{\beta^i, D^j}$ for a rotationally symmetric skyrmion using Usov ansatz $\cos(\theta/2) = (R^2 - r^2)/(R^2 + r^2)$ for $r \leq R$ and 0 for $r > R$. Rows correspond to $\alpha_0$, $\alpha_1$ and $\alpha_2$, expanded in powers of $D/J$. Above, $C \approx 449$. Remaining parameters are given as $\eta_0 = 16\pi/3$, $\eta_D = \eta_0 + (4\pi R/3)(D/J)$.

## TRANSPORT COEFFICIENTS

Texture-independent part of the transport coefficients can be obtained using the Boltzmann equation within the relaxation-time approximation in terms of the integral [1, 2]

$$\mathcal{J}_n^{ij} = \frac{1}{(2\pi)^3 \hbar} \int d\epsilon \tau(\epsilon)(\epsilon - \mu)^n (-\partial_\epsilon f_0) \int dS_\epsilon \frac{v_i v_j}{|\boldsymbol{v}|} \qquad (6)$$

as $\sigma = \mathcal{J}_0$ and $\Pi = -\mathcal{J}_1/\mathcal{J}_0$. Above, $\tau(\epsilon)$ is the relaxation time, $\epsilon(\boldsymbol{k}) = \hbar \omega_k$, $v_i = \partial \omega_k/\partial k_i$, $dS_\epsilon$ is the area $d^2 k$ corresponding to a constant energy surface with $\epsilon(\boldsymbol{k}) = \epsilon$, $f_0$ is the Bose-Einstein equilibrium distribution. Under the assumption that the scattering processes are dominated by Gilbert damping, we set $\tau(\epsilon) \approx 1/2\alpha\omega$. By evaluating the integral after these substitutions, we obtain $\sigma_{\rm 2D} \approx F_{-1}/6\pi^2 \lambda \hbar \alpha$ in three dimensions ($d = 3$), where $\lambda = \sqrt{\hbar J / s k_B T}$ is the wavelength of the thermal magnons, $F_{-1} = \int_0^\infty d\epsilon \epsilon^{d/2} e^{\epsilon+x}/(\epsilon+x)(e^{\epsilon+x} - 1)^2 \sim 1$ evaluated at the magnon gap $x = \hbar \omega_0 / k_B T$. Similarly for $d = 2$, we obtain $\sigma_{\rm 2D} \approx F_{-1}/4\pi \hbar \alpha$.

The spin Seebeck coefficient $L$ is given by $-\hbar \sigma \Pi = \hbar \mathcal{J}_1$, for which we obtain $L_{\rm 3D} \approx F_0 k_B T / 6\pi^2 \lambda \alpha$ in 3D and $L_{\rm 2D} \approx F_0 k_B T / 4\pi \alpha$ in 2D, where $F_0 = \int_0^\infty d\epsilon \epsilon^{d/2}/(e^{\epsilon+x} - 1)^2 \sim 1$. For $d > 2$ and small $x$, the numerical factor $F_0$ can be expressed in terms of Riemann zeta function and Euler gamma function as $\zeta(d/2)\Gamma(d/2 + 1)$ [3]. In the main text, the numerical factors $F_{-1}$ and $F_0$ are omitted.

---



\end{document}